\documentclass[epj]{webofc}
\usepackage[varg]{txfonts}   
\usepackage{graphicx}
\woctitle{The 13th International Symposium on Origin of Matter and Evolution of Galaxies (OMEG 2015)}
\begin{document}
\title{Neutrino-Induced Nucleosynthesis in Helium Shells of Early Core-Collapse Supernovae}
%
%

\author{Projjwal Banerjee\inst{1}\fnsep\thanks{\email{banerjee@physics.umn.edu}}\and
        Yong-Zhong Qian\inst{1,5}\and
        Alexander Heger\inst{2,5}\and
        Wick Haxton\inst{3,4}
}

\institute{School of Physics and Astronomy, University of Minnesota, Minneapolis, Minnesota 55455, USA
\and
           Monash Center for Astrophysics, School of Physics and Astronomy, Monash University, Victoria 3800, Australia
\and
           Department of Physics, University of California Berkeley, Berkeley, California 94620, USA
\and
           Lawrence Berkeley National Laboratory, Berkeley, California 94620, USA
\and
          Center for Nuclear Astrophysics, INPAC, Department of Physics and Astronomy, Shanghai Jiao Tong University, Shanghai 200240, People’s Republic of China
          }

\abstract{%
We summarize our studies on neutrino-driven nucleosynthesis in He shells of 
early core-collapse supernovae with metallicities of $Z\lesssim 10^{-3}\,Z_\odot$. 
We find that for progenitors of $\sim 11$--$15\,{\rm M}_\odot$, the neutrons released
by $^4{\rm He}(\bar{\nu}_e,e^+n)^3{\rm H}$ in He shells can be captured to 
produce nuclei with mass numbers up to $A \sim 200$. This mechanism is sensitive 
to neutrino emission spectra and flavor oscillations. In addition, we find two new 
primary mechanisms for neutrino-induced production of $^{9}$Be in He shells. 
The first mechanism produces $^9$Be via 
$^7{\rm Li}(n,\gamma)^8{\rm Li}(n,\gamma)^9{\rm Li}(e^-\bar{\nu}_e)^9{\rm Be}$ 
and relies on a low explosion energy for its survival. The second mechanism 
operates in progenitors of $\sim 8\,{\rm M}_\odot$, where $^9$Be can be produced directly 
via $^7{\rm Li}(^3{\rm H},n_0)^9{\rm Be}$ during the rapid expansion of the shocked 
He-shell material. The light nuclei $^7$Li and $^3$H involved in these mechanisms 
are produced by neutrino interactions with $^4$He. We discuss the implications of 
neutrino-induced nucleosynthesis in He shells for interpreting the elemental
abundances in metal-poor stars.}
\maketitle
\section{Introduction}
\label{intro}

Observations of the chemical composition in metal-poor stars provide important 
clues for the source of metal enrichment in the early Galaxy. In this regard, 
extremely metal-poor (EMP) stars with 
${\rm [Fe/H] \equiv log(Fe/H)-log(Fe/H)_\odot \lesssim -2.5}$ are especially 
useful as they sample gases polluted by just a few nucleosynthetic events. 
With a growing number of observational studies on EMP stars, it has now 
become clear that neutron-capture elements were present in the early Galaxy, 
where only stars more massive than $\sim 8\,{\rm M}_\odot$ that end their 
lives as core-collapse supernovae (CCSNe) could have contributed \cite{mp_obs}.  
The abundance patterns observed in EMP stars show large variations. 
Some stars have abundance patterns similar to the solar rapid  
neutron-capture ($r$) process distribution and others have patterns similar to the slow
neutron-capture ($s$) process distribution 
or to a mixture of $r$ and $s$ process distributions \cite{beers2005}. 
This suggests a variety of neutron-capture processes operated in the early Galaxy. EMP
stars with $s$-process signatures are thought to have experienced surface pollution 
by a binary companion during the latter's evolution on the asymptotic giant branch. 
However, the astrophysical site of $r$-process in the early Galaxy and 
the mechanism for producing a mixture of $r$ and $s$ patterns, especially in stars 
that are not in binary systems, are still uncertain. We revisit a mechanism for 
producing neutron-capture elements in the early Galaxy originally proposed by 
Epstein, Colgate, and Haxton  (ECH) \cite{ech}. We show that in metal-poor 
progenitors with metallicities of $Z\lesssim 10^{-3}\,Z_\odot$ and masses of
$\sim 11$--$15\,{\rm M}_\odot$, a modified version of the ECH mechanism can lead to 
production of nuclei with mass numbers up to $A\sim 200$ \cite{bhq}. 
We identify a related mechanism for producing the rare light nucleus
$^{9}$Be that previously had been thought to be produced exclusively by interactions between 
Galactic comic rays (GCRs) and nuclei in the interstellar medium \cite{ramaty,prantzos}. 
Furthermore,  $^{9}$Be can be synthesized by neutrino-induced 
reactions in He shells of more compact progenitors of $\sim 8$--$10\,{\rm M}_\odot$. 
We find that $^9$Be produced by the above mechanisms can account for the  
linear relationship of Be with primary products of CCSNe, e.g., O, Mg, Ti, and Fe, 
as observed at low metallicities \cite{boesgaard}.

\begin{figure}
\centering
\includegraphics[width=0.45\textwidth]{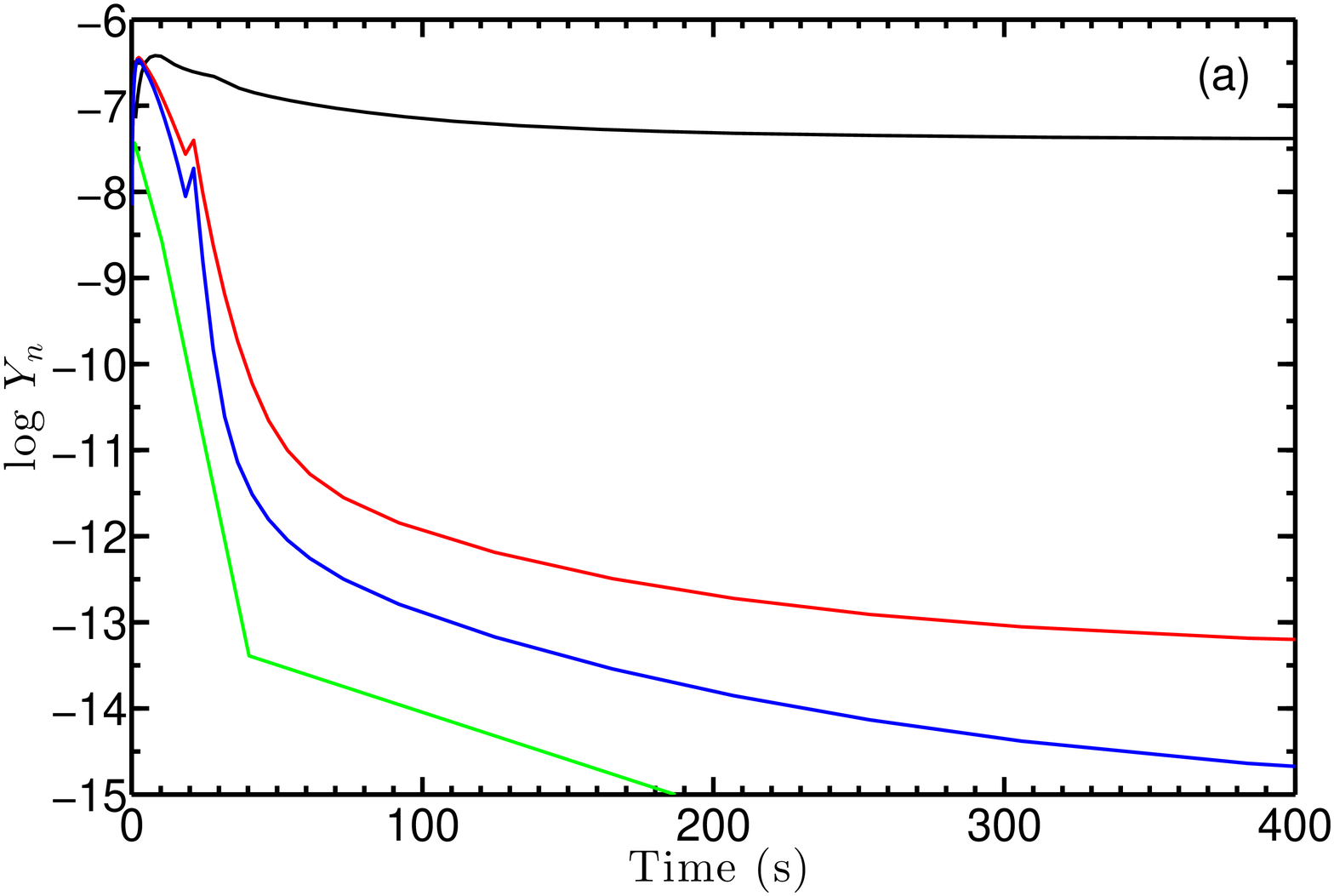}
\includegraphics[width=0.45\textwidth]{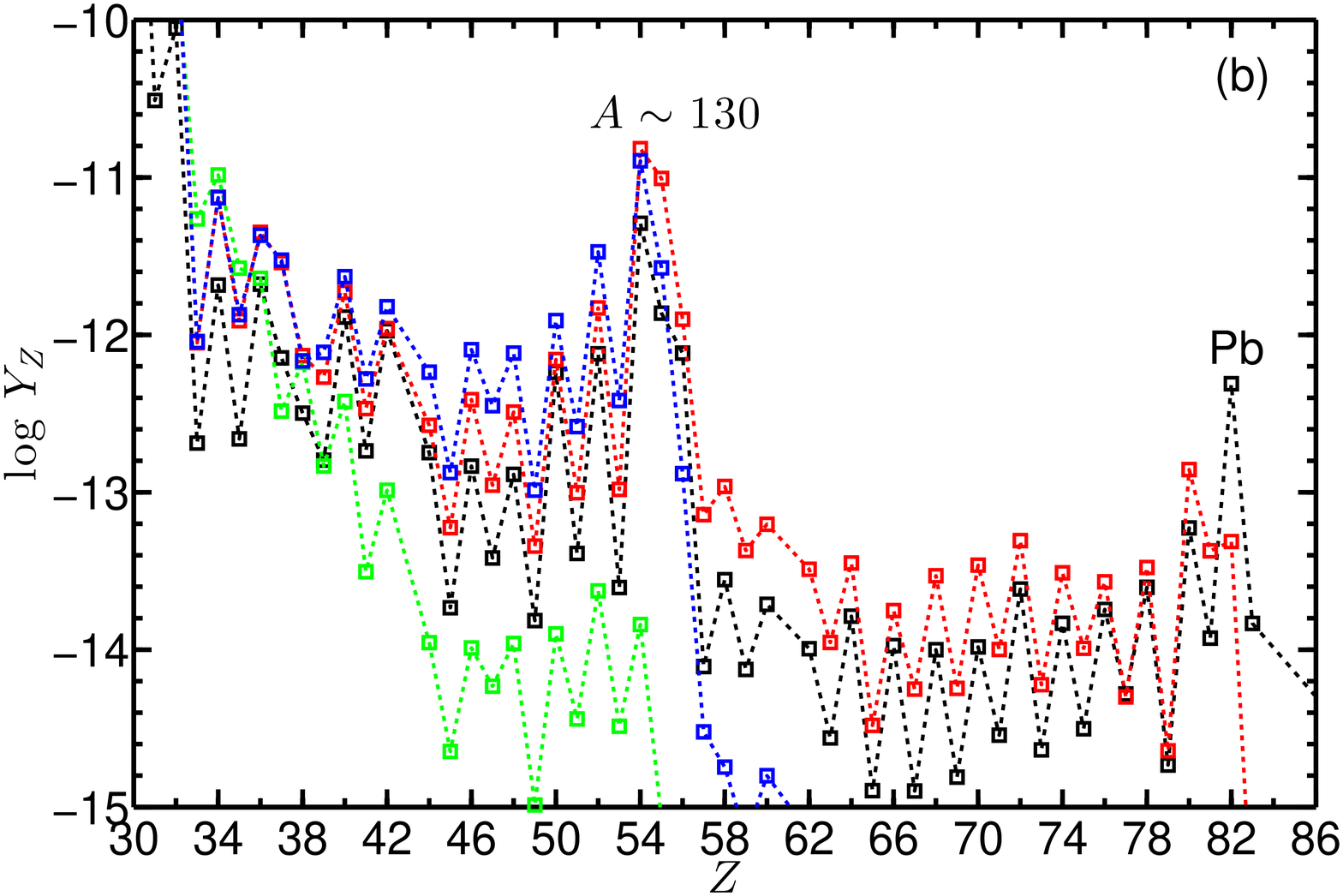}
\label{fig-ncomp}
\caption{(a) Evolution of neutron abundance $Y_n$ with time for a typical zone 
in the outer He shell for 
y11$\overline{\rm H}.1$ (\emph{black}), u11*$\overline{\rm H}.1$ (\emph{red}), 
u11$\overline{\rm H}.1$ (\emph{blue}), and v11$\overline{\rm H}.1$ (\emph{green}).
(b) Final elemental abundance pattern produced by 
y11$\overline{\rm H}.1$ (\emph{black}), u11*$\overline{\rm H}.1$ (\emph{red}), 
u11$\overline{\rm H}.1$ (\emph{blue}), and v11$\overline{\rm H}.1$ (\emph{green}).}
\vspace*{-0.7cm}
\end{figure}

\section{Neutrino-Induced Neutron-Capture Process in the He Shell}
The ECH mechanism was proposed as a possible scenario for the $r$ process in CCSNe.  
The main idea is that neutrinos emitted during a CCSN
will produce neutrons in neutral-current 
reactions with $^4$He via 
$^4{\rm He}(\nu,\nu' n)^3{\rm He}(n,p)^3{\rm H}(^3{\rm H},2n)^4{\rm He}$ and
$^4{\rm He}(\nu,\nu' p)^3{\rm H}(^3{\rm H},2n)^4{\rm He}$. 
As neutrons cannot capture on $^4$He, they will instead be absorbed with high efficiency 
by trace quantities of $^{56}$Fe present from initial birth material of the star.
As the number 
of neutrons generated by this primary neutrino-driven mechanism can be computed reasonably 
well, it can be shown that the neutron/seed ratio needed for a successful 
$r$ process can be achieved in metal-poor He shells.
It was pointed out in Ref.~\cite{WHHH90} however, that this mechanism would only be 
viable in CCSNe from low-metallicity compact stars and even there poisons such as 
$^{12}$C may inhibit the $r$ process. We revisit this scenario by performing detailed 
nucleosynthesis calculations with the hydrodynamic code KEPLER \cite{kepler1,kepler2}. 
We use the same code to evolve progenitor models and treat the effects of the CCSN
shock. Progenitors investigated have metallicities of $[Z]\equiv\log(Z/Z_\odot)=-5$ 
(denoted by ``y'' for ``hyper-metal-poor''), $-4$ (``u'' for ``ultra-metal-poor''), and $-3$ 
(``v'' for ``very-metal-poor'') and masses of $11$--$15\,{\rm M}_\odot$. The progenitor
models are labelled by their metallicity and mass in units of ${\rm M}_\odot$.
For example, u11 refers to a progenitor with metallicity ``u'' of $11~{\rm M}_\odot$. 
We study the sensitivity of the final abundance pattern to neutrino spectra, flavor 
oscillations, explosion energy, and progenitor metallicity. 

The progenitor models y11, y15, u11, u15, v11, and v15 are generated using the most 
recent version of KEPLER with a full reaction network to track the evolution of a massive 
star from birth till right before core collapse. We also consider modified models u11* and 
u15*, which have the abundances of $^{28}$S and $^{32}$Si in the He shell reduced by 
a factor of $\sim 10$ to match similar models in Ref.~\cite{WHW}.
The explosion is modeled by driving a piston through the collapsing progenitor and then
following the propagation of the shock wave \cite{kepler1}. The nucleosynthesis due to 
shock heating and that due to neutrino emission from the protoneutron star (PNS) are 
followed using the full reaction network. The neutrino luminosity is taken to be 
$L_\nu(t)=L_\nu(0)\exp(-t/\tau_\nu)$ for each of the six species,  
with $L_\nu(0)=16.7$~B/s ($1\ {\rm B} \equiv 10^{51}$~erg) and $\tau_\nu=3$~s, 
which amounts to a total energy of 300~B emitted in neutrinos. We adopt Fermi-Dirac
neutrino spectra with zero chemical potential and constant temperatures
$T_{\nu_e}$, $T_{\bar\nu_e}$, and $T_{\nu_x}=T_{\bar\nu_x}$ ($x =\mu,\tau$).
We take $(T_{\nu_e},T_{\bar\nu_e},T_{\nu_x})=(4, 5.33, 8)$~MeV (H) and
$(3, 4, 6)$~MeV (S), which represent the harder and softer spectra
obtained from earlier (e.g., \cite{woosley}) and more recent
\cite{hudepohl,fischer} neutrino transport calculations, respectively. 

Similar to Ref.~\cite{bhq}, the site for the neutrino-induced neutron-capture process is 
the outer He shell where the composition is nearly pure $^4$He and mass fractions 
of poisons such as $^{12}$C, $^{14}$N, and $^{16}$O are so low that they capture 
very few neutrons. For the models considered here, this corresponds to a 
typical radius of $\sim 10^{10}$~cm, a pre-shock temperature of $\sim 9\times10^7$~K 
and density of $\sim 50\ {\rm g/cm}^3$.
We calculate neutrino-induced nucleosynthesis both before and after the arrival of the shock.
The neutrino spectra in the outer He shell should differ from the 
initial spectra at the PNS surface due to flavor oscillations associated with the atmospheric
neutrino mass splitting  $|\delta m_{13}^2| \sim 2.4\cdot 10^{-3}$~eV$^2$ 
(see e.g., \cite{nu_osc1,nu_osc2}): the neutrinos experience a level crossing before they 
reach the He shell. For the inverted neutrino mass hierarchy (IH), 
this results in almost full $\bar\nu_e \leftrightarrow \bar\nu_x$ conversion and consequently
an enhanced charged-current rate for $^4{\rm He}(\bar\nu_e,e^+n)^3{\rm H}$.
This resulting increase in neutron production is helpful in achieving
the necessary neutron/seed ratio \cite{bhq}. 
In contrast, for the normal neutrino mass hierarchy (NH), 
the rate of $^4{\rm He}(\nu_e,e^-p)^3{\rm He}$ is increased, producing more $^3$He 
to dominate neutron capture, making an $r$-process impossible. 
So the NH precludes an efficient neutron-capture process. 
We explore the case of full $\bar\nu_e \leftrightarrow \bar\nu_x$
interconversion as well as the case where flavor oscillations are ignored. 
We also explore a range of explosion energies
$E_{\rm expl}=0.1$--1~B. The nucleosynthesis calculations are labelled by the 
progenitor model, the neutrino spectra, and the explosion energy in units of B, with e.g., 
u8.1H.1 indicating progenitor model u8.1, the harder neutrino spectra H, and 
$E_{\rm expl} = 0.1$~B.  Calculations including $\bar\nu_e \leftrightarrow \bar\nu_x$ 
oscillations are denoted by a bar above the H or S.  

We find that the highest He-shell neutron abundance and thus the 
most favorable conditions for an efficient neutron-capture process are obtained with 
a hard spectra with oscillations for the IH. The neutron abundances $Y_n$ for $11\,{\rm M}_\odot$ progenitors of different metallicities are shown in Fig.~\ref{fig-ncomp}a.
We find that the neutral-current reactions $^4{\rm He}(\nu,\nu' n)^3{\rm He}$ are ineffective in 
producing neutrons as they are immediately recaptured via $^3{\rm He}(n,p)^3{\rm H}$. 
In addition, the potential neutron-producing reaction $^3{\rm H}(^3{\rm H},2n)^4{\rm He}$ 
is ineffective at the temperatures in the outer He shell. Instead, 
neutrons are primarily produced by the charged-current reaction 
$^4{\rm He}(\bar\nu_e,e^+n)^3{\rm H}$. While this production 
occurs over the $\sim 5$--6~s period, the neutron-capture process takes place 
over $\sim 40$--500~s, producing nuclei up to $A\sim 200$. Both
$^7$Li and $^8$Li are produced \textsl{in situ} as a result of neutrino reactions
and act as important poisons. The nucleus $^7$Li is produced 
by $^4{\rm He}(^3{\rm H},\gamma)^7{\rm Li}$, 
where $^3$H is made by $^4{\rm He}(\bar\nu_e,e^+n)^3{\rm H}$ and 
$^4{\rm He}(\nu,\nu' p)^3{\rm H}$. The nucleus $^8$Li is produced by 
neutron capture on $^7$Li. Enough $^9$Be is produced via 
$^7{\rm Li}(n,\gamma)^8{\rm Li}(n,\gamma)^9{\rm Li}(e^-\bar\nu_e)^9{\rm Be}$ to
account for the $^9$Be observed in metal-poor stars \cite{bqhh}. 
This is discussed in more detail in the next section. 

For the y11$\overline{\rm H}.1$ model, neutrons are never depleted, which 
allows neutron capture to last for $\sim 500$~s before the neutron
density becomes too low. Such a long timescale causes a considerable amount 
of Pb to be formed well after 100~s. 
In addition, availability of high neutron density for $\sim 500$~s allows 
capture on seeds such as Si, S, Ca, and Ti to contribute to the
production of nuclei up to $A\gtrsim 130$. This is unique to 
the y11$\overline{\rm H}.1$ model and is not seen in other models where 
neutrons are depleted within $\lesssim 100$~s.
For the u* models, hard spectra with $\bar\nu_e \leftrightarrow \bar\nu_x$ oscillations 
are required for the nuclear flow to reach $A\sim 200$ (Fig.~\ref{fig-ncomp}b) but 
hard spectra without oscillations can reach $A\sim 130$.
For the u models ($[Z]=-4$), $^{28}$Si and $^{32}$S start to become important poisons, 
competing with $^{56}$Fe for the neutrons. 
In this case, the flow can still reach $A \sim 130$ for hard neutrino spectra with 
$\bar\nu_e \leftrightarrow \bar\nu_x$ oscillations. As can be seen from 
Fig.~\ref{fig-ncomp}b, neutron capture starts to shut off for the v11 and v15 models 
when the metallicity reaches $[Z]=-3$.

\begin{table}
\centering
\caption{Yields of Be, Sr, and Ba in ${\rm M}_\odot$ [$X(Y) \equiv X\cdot10^Y$]
and the corresponding [Sr/Ba] with $\log{\rm (Sr/Ba)}_\odot=0.70$} 
\label{tab-1}       
\begin{tabular}{lrrrrr}
\hline
 Model                   &Be           &Sr                 &Ba              &[Sr/Ba]\\
\hline
y11$\overline{\rm H}.1$  & $1.48(-8)$    &   $3.02(-10)$  & $1.14(-9)$     &$-1.09$\\
y11$\overline{\rm H}.3$  & $7.90(-9)$    &   $2.35(-10)$  & $3.21(-10)$    & $-0.65$\\
y15$\overline{\rm H}.1$  & $1.07(-8)$    &  $3.43(-11)$   & $1.05(-10)$    & $-1.00$\\
y15$\overline{\rm H}.3$  & $1.01(-9)$    &  $6.91(-11)$   & $1.40(-10)$    & $-0.82$\\

u11*$\overline{\rm H}.1$ & $1.20(-8)$    &  $6.50(-10)$  & $1.87(-9)$    &$-0.97$ \\
u11*$\overline{\rm H}.3$ & $2.48(-9)$    &  $5.70(-10)$  & $1.90(-9)$    &$-1.03$ \\
u15*$\overline{\rm H}.1$ & $2.44(-9)$    &  $1.32(-10)$   & $4.86(-11)$   & $-0.08$\\
u15*$\overline{\rm H}.3$ & $8.46(-10)$    &  $1.22(-10)$   & $9.50(-11)$  & $-0.40$\\

u11$\overline{\rm H}.1 $ & $2.72(-9)$    &   $5.99(-10)$  & $1.95(-10)$   & $-0.02$\\
u11$\overline{\rm H}.3 $ & $7.13(-10)$   &  $4.92(-10)$  & $1.84(-10)$   & $-0.09$\\
u15$\overline{\rm H}.1$  & $9.92(-10)$   &   $2.18(-10)$  & $3.27(-12)$  & 1.31\\
u15$\overline{\rm H}.3$  & $2.23(-10)$   &   $3.34(-10)$  & $7.33(-12)$  & 1.14\\
\hline
\end{tabular}
\vspace*{-0.5cm}
\end{table}

\section{Neutrino-Induced Production of $^{9}$Be in the He Shell}
The neutrino-induced neutron-capture process discussed above can also 
lead to $^{9}$Be synthesis \cite{bqhh}. In addition to neutron capture mainly
by $^{56}$Fe leading to production of nuclei up to $A\sim 200$, $^{9}$Be is 
made by a ``mini-\textsl{r} process" through
$^7{\rm Li}(n,\gamma)^8{\rm Li}(n,\gamma)^9{\rm Li}(e^-\bar\nu_e)^9{\rm Be}$. 
In this case, $^7$Li and $^8$Li act as poisons for the main neutron-capture
process but serve as the seeds for $^{9}$Be production. 
Most of the $^9$Be is produced before the arrival of the shock at 
temperatures $T\lesssim 10^8$ K where the newly-synthesized $^9$Be 
survives. The $^{9}$Be production ceases at shock arrival, which produces temeratures 
$\gtrsim 2\times 10^8$ K where $^9{\rm Be}(p,{^4{\rm He}})^6{\rm Li}$ and $^9$Be$(p,d)2{^4{\rm He}}$ can destroy part of the $^9$Be. 
Thus the final $^{9}$Be yield is sensitive to the shock temperature, and hence,
the explosion energy. For $E_{\rm expl} \gtrsim3$~B, most of the $^{9}$Be is 
destroyed. In addition, as in the main neutron-capture process, 
$^{9}$Be production is sensitive to the neutrino spectra, becoming most efficient for hard 
spectra with oscillations ($\overline{\rm H}$ models). However, as only two 
neutrons need to be captured, models with hard spectra without oscillations 
(H models) and soft spectra with oscillations ($\overline{\rm S}$ models) are 
also able to produce substantial amounts of $^{9}$Be. As in the 
neutrino-induced neutron-capture process, $^{9}$Be production in the He shell ceases for metallicities of $[Z]\gtrsim -3$.
   
\begin{figure}
\centering
\includegraphics[width=8cm,clip]{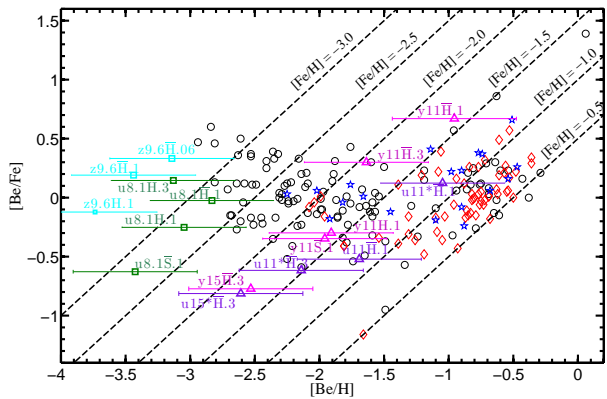}
\caption{Comparison of [Be/H] and [Be/Fe] from models with data 
from \cite{boesgaard} (\emph{black circles}), \cite{smiljanic}  (\emph{red diamonds}), 
and \cite{tan} (\emph{blue stars}). The z models correspond to progenitors with zero
metallicity.}
\label{fig-Be}
\vspace*{-0.7cm}
\end{figure}

There is a second He-shell neutrino-driven 
mechanism for producing $^9$Be that operates in progenitors with masses of 
$\sim 8$--$9.6\,{\rm M}_\odot $, which are near the lower limit for producing a CCSN. Such 
progenitors are compact, with a very thin He shell and a 
steeply-falling density profile outside the core. The structure is quite 
different from that found in progenitors of 
$\gtrsim 11\,{\rm M}_\odot$: the He shell of a progenitor of $\sim 8\,{\rm M}_\odot$
is located at a radius of $\sim 10^9$ cm, ten times closer to the core than in an  
$11\,{\rm M}_\odot$ progenitor. As a result, the shock reaches the He shell within $\sim 1$~s. 
The peak post-shock temperature reaches $\sim 8\times 10^8$~K, which destroys all 
of the $^{9}$Be produced before the arrival of the shock. Due to the compact 
structure however, as the shocked He-shell material expands, the temperature drops below 
$2\times 10^8$~K within $\sim 4$~s and destruction of $^9$Be via 
$^9$Be$(p,{^4{\rm He}})^6{\rm Li}$ and $^9$Be$(p,d)2{^4{\rm He}}$ is effectively 
turned off. At this stage, the He shell is still close to the PNS and is still irradiated by 
neutrinos, which results in the breakup of $^4$He via $^4{\rm He}(\nu,\nu' n)^3{\rm He}$, 
$^4{\rm He}(\nu,\nu' p)^3{\rm H}$, and $^4{\rm He}(\bar\nu_e,e^+ n)^3{\rm H}$. 
Then $^9$Be is made from the breakup products via 
$^4{\rm He}(^3{\rm H},\gamma)^7{\rm Li}(^3{\rm H},n_0)^9{\rm Be}$. It is 
important to note that the above production of $^9$Be also happens in more massive 
progenitors of $\gtrsim 11\,{\rm M}_\odot$, but in this case all the $^9$Be produced is 
destroyed by the shock. Due to the slower expansion in these progenitors, by the time 
the temperature in the inner He shell drops below $2\times 10^8$~K at $\sim 25$~s, 
neutrino emission by the PNS has effectively ceased. Thus the fast expansion
due to the sharp density gradient in low-mass progenitors is essential to the production 
of $^9$Be via this mechanism. This mechanism is also sensitive to the neutrino spectra
due to the high thresholds for the breakup 
reactions $^4{\rm He}(\nu,\nu' p)^3{\rm H}$ and $^4{\rm He}(\bar\nu_e,e^+ n)^3{\rm H}$. 
For example, the $^9$Be yield for u8.1$\overline{\rm H}.1$ is $\sim4$ times higher than
for u8.1$\overline{\rm S}.1$. 

In Fig.~\ref{fig-Be} we compare model yields of [Be/H] and [Be/Fe] with the data. 
The horizontal bar corresponds to the estimated range for the amount
of the interstellar medium to mix with the ejecta from each CCSN for the relevant 
explosion energy \cite{thornton}. It can be seen that the predicted [Be/H] and [Be/Fe]
values from most of the models with hard neutrino spectra agree very well with 
observations and can account for the observed Be up to [Fe/H]~$\sim -1$. 
For the lower explosion energies, some inner ejecta from the 11--$15\,{\rm M}_\odot$ models
is likely to fall back onto the PNS so that only a fraction of the Fe produced is 
ejected while all of the He shell containing the $^9$Be produced by the first mechanism
is ejected. In this case, the predicted [Be/Fe] values
in Fig.~\ref{fig-Be} refer to the lower limits for such models. For the second mechanism 
of $^9$Be production that operates in lower-mass CCSNe, all the material is ejected
even for low explosion energies. It is possible that the first mechanism operating in
11--$15\,{\rm M}_\odot$ CCSNe can contribute to the Be in stars with [Fe/H]~$\sim -2$ to 
$-1$, whereas the second mechanism operating in lower-mass CCSNe can account 
for the Be in stars with [Fe/H]~$\lesssim -2.5$. 

\section{Discussion and Conclusions}
We have studied neutrino-induced nucleosynthesis in He shells of CCSNe resulting from 
metal-poor progenitors of both lower masses of $\sim 8$--$9.6\,{\rm M}_\odot$ with compact 
structure and regular masses of 11--$15\,{\rm M}_\odot$. In regular-mass CCSNe, neutrino 
interactions with $^4$He can produce neutrons, which are then captured by Fe seeds to
produce heavy nuclei up to $A\sim 200$. Compared to the usual $r$ process, the 
neutrino-induced neutron-capture process in He shells of metal-poor CCSNe is cold (temperatures of $\sim 10^8$~K)  and slow (timescales ranging to several hundred 
seconds). This mechanism is sensitive to neutrino spectra, flavor oscillations, and 
metallicity of the progenitor. Hard spectra with oscillations are required for most models 
to produce nuclei at and beyond Ba. Depending on the metallicity and neutrino spectra, 
this mechanism can account for a wide range of [Sr/Ba] from $\sim -1$ to 1.3 (see 
Table.~\ref{tab-1}). This is consistent with the large scatter for [Sr/Ba] observed
in metal-poor stars of [Fe/H]~$\lesssim -2.5$. We find that for typical Sr yields of 
$\sim (3$--$8)\times 10^{-10}\,{\rm M}_\odot$ mixed with $\sim 10^2$-$10^3\,{\rm M}_\odot$ of the 
interstellar medium appropriate for weak explosions \cite{thornton}, various models 
can account for the [Sr/H] and [Sr/Ba] observed in many of the metal-poor stars 
with low abundances of neutron-capture elements.

The neutrino-induced neutron-capture process in low-metallicity 
He shells of 11--$15\,{\rm M}_\odot$ 
CCSNe also results in $^9$Be synthesis before the arrival 
of the shock, which can account for the Be observed in metal-poor stars. 
In addition, low-mass compact progenitors of $\sim 8$--$10\,{\rm M}_\odot$ can 
produce $^9$Be in the shocked fast-expanding He shell due to the continuing neutrino 
irradiation. Both of these mechanisms are consistent with the observed primary 
nature of Be at low metallicities. Other mechanisms, such as GCR interactions
with the interstellar medium, are still required to account for the Be observed at 
higher metallicities including the solar inventory of Be.

This work was supported in part by the US DOE [DE-FG02-87ER40328 (UM), 
DE-SC00046548 (Berkeley), and DE-AC02-98CH10886 (LBL)] 
and by ARC Future Fellowship FT120100363 (AH).


%
%
%

\bibliography{OMEG_banerjee}
\bibliographystyle{woc}

\end{document}